\documentclass[3p]{elsarticle}

\usepackage{lineno,hyperref}
\modulolinenumbers[5]
\usepackage{amsmath} 
\usepackage{amssymb}
\usepackage{units} 			
\usepackage{tabularx}   
\usepackage{multirow}  
\usepackage{threeparttable}  
\newcommand{\bu}{{\boldsymbol u}}
\newcommand{\bx}{{\boldsymbol x}}
\newcommand{\parci}[2]{\frac{\partial #1}{\partial #2}}

\begin{document}

\begin{frontmatter}

\title{Revisiting the single-phase flow model for liquid steel ladle stirred by gas}

\author[mymainaddress]{Najib Alia\corref{mycorrespondingauthor}}
\cortext[mycorrespondingauthor]{Corresponding author}
\ead{alia@wias-berlin.de}

\author[mymainaddress,mysecondaryaddress]{Volker John}
\ead{john@wias-berlin.de}

\author[mythirdaddress]{Seppo Ollila}
\ead{seppo.ollila@ssab.com}

\address[mymainaddress]{Weierstrass Institute for Applied Analysis and Stochastics (WIAS), Mohrenstra{\ss}e 39, D-10117 Berlin, Germany}
\address[mysecondaryaddress]{Freie Universit\"at, Department of Mathematics and Computer Science, Arnimallee 6, D-14195 Berlin, Germany}
\address[mythirdaddress]{SSAB Europe Oy, Rautaruukintie 155, FI-92101 Raahe, Finland}

\begin{abstract}
Ladle stirring is an important step of the steelmaking process to homogenize the temperature and the chemical composition of the liquid steel and to remove inclusions before casting. 
Gas is injected from the bottom of the bath to induce a turbulent flow of the liquid steel. 
Multiphase modeling of ladle stirring can become computationally expensive, especially 
when used within optimal flow control problems. This paper focuses therefore on single-phase flow models.
It aims at improving the existing models from the literature.
Simulations in a 2d axial-symmetrical configuration, as well as, in a real 3d laboratory-scale ladle, are 
performed.
The results obtained with the present model are in a relative good agreement with experimental data and suggest that it can be used as an efficient model in optimal flow control problems.
\end{abstract}


\end{frontmatter}


\section{Introduction}\label{sec:1}

\begin{figure}[t!]
	\begin{center}
		\includegraphics[width=0.25\textwidth]{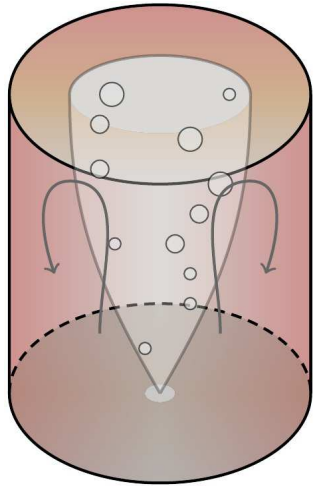}
	\end{center}
	\caption{Schematic sketch of ladle stirring.}	\label{fig:3dcylinders}
\end{figure}

Gas stirring in ladles is a standard practice in the steelmaking industry to refine and homogenize the liquid steel bath before casting. In this process, a noble gas is injected from the bottom of the ladle through generally one or two nozzles, called porous plugs. 
The gas rises by buoyancy through the liquid steel, forms a gas plume, and causes stirring, i.e., a mixing of the bath (Figure~\ref{fig:3dcylinders}).

The final aim of our research consists in developing an approach for obtaining an optimal homogenization
of the liquid steel by controlling the flow rate of the injected gas, based on a mathematical model. Thus, 
an optimal flow control problem has to be solved. The solution of such a problem requires, on the one hand, 
the repeated solution of the considered process with slightly changing coefficients. If gradient-based 
optimization algorithms shall be used, note that gradient-free schemes converge generally very slowly, 
one has to compute the derivative of the objective functional with 
respect to the solution of the partial differential equation (PDE). This step can be performed most efficiently 
by solving an adjoint equation. For time-dependent problems, the adjoint equation is backward in time and 
for nonlinear problems, the coefficients of the adjoint equation depend on the computed solution of the PDE. 
In the case of ladle stirring, one has to consider a time-dependent nonlinear model. 
Altogether, on the other hand, one has to solve repeatedly an adjoint problem with slightly changing coefficients. 

To keep the cost of the optimal control problem reasonable, reduced models are often 
applied. The results obtained with these models are computed much faster and they are less accurate than with the full model. But, usually, 
the accuracy is sufficient for the optimal control problem. There are several approaches for defining reduced models. 
A modern one, called reduced order modeling (ROM),  consists in computing bases for the discrete problems that
already possess information about 
the solution. For time-dependent problems, one performs one simulation with the full model with 
certain coefficients, stores the so-called snapshots, and computes a basis via a proper orthogonal 
decomposition, see \cite{Sir87,NMT11,CIJS14} for detailed descriptions. However, this approach becomes complicated for complex
mathematical models and their adjoint equations. In addition, it is generally only available in some academic 
research software. If commercial software is utilized, like in the present paper, one has usually only the 
option to consider from the beginning a mathematical model that is sufficiently efficient. For this reason, single-phase 
models, which are based on the standard mono-phase incompressible Navier--Stokes equations only (\cite{Mazumdar1995}), will be studied here, 
instead of two-phase gas-liquid flow models. The goal of this paper consists in defining an appropriate model which can be used in the simulation 
of optimal flow control problems.

In the literature, the so-called ``quasi-single phase models'' are based on the main assumption that the gas fraction $\alpha$ in the fluid domain is known.
Debroy et al.~\cite{Debroy1978}, Gr\'{e}vet et al.~\cite{Grevet1982}, and Sahai and Guthrie~\cite{Sahai1982} were among the firsts to perform such numerical simulations on gas-stirred cylindrical ladles with one central nozzle. They applied a conical plume geometry and three slightly different formulas for $\alpha$. In~\cite{Sahai1982}, the model additionally uses a moving wall boundary condition, e.g., a non-homogeneous Dirichlet condition on the central vertical axis. Later, Balaji and Mazumdar~\cite{Balaji1991} adapted the existing formulas to propose
a fourth variant of $\alpha$, and obtained a better agreement of the numerical results with experimental measurements. These findings were summarized by Mazumdar and Evans in~\cite{Mazumdar2009}. In parallel to these works, Woo et al.~\cite{Woo90} developed empirical formulas for $\alpha$ based on experimental measurements and obtained better numerical results than in \cite{Debroy1978} and~\cite{Grevet1982}.
Single-phase modeling for bubbles columns was also applied for chemical applications by  Bernard et al.~\cite{Bernard2000}. The authors applied a formula similar to~\cite{Sahai1982}, except that they included a height correction factor to take into account the volumetric expansion of the rising gas bubbles.

All these single-phase models found a practical application in several works, such as, for example, the numerical study of mass and heat transfer phenomena (\cite{Mazumdar1992,Ganguly2004,Mukhopadhyay2001}), the improvement of stirring by changing ladle geometry and nozzle positions (\cite{Zhu96,Goldschmit2001}), or the comparison with two-phase flow models (\cite{Mazumdar1994,Mazumdar1995b}).
Table~\ref{table:review} summarizes the literature dealing with `quasi-single phase' models.

\begin{table}
	\begin{center}
		\caption{Summary of the studies based on `quasi-single phase' models.}
			\begin{threeparttable}
			\begin{tabular}{   *{1}{>{\centering\arraybackslash}m{0.5cm}} *{1}{>{\centering\arraybackslash}m{1.5cm}} *{1}{>{\centering\arraybackslash}m{2.5cm}} *{1}{>{\centering\arraybackslash}m{8.cm}} }
				\noalign{\smallskip}\hline\noalign{\smallskip}
				Ref.	& Year  & Model based on  & Object of the study	 \\ \noalign{\smallskip}\hline\noalign{\smallskip}
				\cite{Mazumdar1995} & 1995 & \cite{Debroy1978,Grevet1982,Sahai1982,Balaji1991,Woo90}\tnote{*} & Review of existing `quasi-single phase' models \\ \noalign{\smallskip}  \noalign{\smallskip}
				\cite{Debroy1978} & 1978 & \cite{Debroy1978} & Definition for $\alpha$ and application in axial-symmetrical ladles \\ \noalign{\smallskip} \noalign{\smallskip}
				\cite{Grevet1982} & 1982 & \cite{Grevet1982}  & New definition for $\alpha$ \\ \noalign{\smallskip} \noalign{\smallskip}
				\cite{Sahai1982} & 1982 & \cite{Sahai1982} & New definition for $\alpha$ in combination with vertical boundary velocity  \\ \noalign{\smallskip} \noalign{\smallskip}
				\cite{Balaji1991} & 1991 & \cite{Balaji1991,Grevet1982,Sahai1982} & New definition for $\alpha$ and comparison with older versions   \\ \noalign{\smallskip} \noalign{\smallskip}
				\cite{Mazumdar2009} & 2010 & \cite{Balaji1991} & -  \\ \noalign{\smallskip} \noalign{\smallskip}
				\cite{Woo90} & 1990 & \cite{Woo90,Debroy1978,Sahai1982} & New definition for $\alpha$ and comparison with older versions \\ \noalign{\smallskip} \noalign{\smallskip}
				\cite{Mazumdar1992} & 1992 & \cite{Balaji1991}\tnote{*} & Application on mass transfer rates \\ \noalign{\smallskip} \noalign{\smallskip}
				\cite{Ganguly2004} & 2004 & \cite{Sahai1982}\tnote{*} & Study of thermal stratification    \\ \noalign{\smallskip} \noalign{\smallskip}
				\cite{Mukhopadhyay2001} & 2001 & \cite{Woo90}\tnote{*} & Application on temperature and heat transfer \\ \noalign{\smallskip} \noalign{\smallskip}
				\cite{Zhu96} & 1996 & \cite{Woo90} & Effect of geometry and nozzle position on mixing time  \\ \noalign{\smallskip} \noalign{\smallskip}
				\cite{Goldschmit2001} & 2001 & \cite{Balaji1991} & Effect of geometry and nozzle position on the circulation rate \\ \noalign{\smallskip}  \noalign{\smallskip}
				\cite{Mazumdar1994} & 1994 & \cite{Balaji1991}\tnote{*} & Comparison with Euler-Euler and Euler-Lagrange models \\ \noalign{\smallskip}  \noalign{\smallskip}
				\cite{Mazumdar1995b} & 1995 & \cite{Balaji1991}\tnote{*} & Comparison with Euler-Euler models  \\ \noalign{\smallskip}  \hline \noalign{\smallskip}
		\end{tabular}
	\begin{tablenotes}
		\item[*] With minor changes in comparison to original reference.
	\end{tablenotes}
		\end{threeparttable}
		\label{table:review}
	\end{center}
\end{table}

This paper proposes to re-visit the single-phase model for ladle stirring with the following contributions:
\begin{itemize}
	\item clarify the definitions and differences between the most often used formulas for $\alpha$ (\cite{Sahai1982,Balaji1991,Woo90,Mukhopadhyay2001}),
	\item simplify the model, so that standard incompressible Navier--Stokes solvers can be used,
	\item validate the model with existing results and determine the most appropriate gas fraction $\alpha$ for 
	a 2d axial-symmetrical configuration,
	\item and validate the model in 3d on a recent laboratory-scale water ladle experiment~\cite{Palovaara2018}.
\end{itemize}
All simulations were performed with the commercial software {\sc Comsol Multiphysics\textsuperscript{\textcopyright}}.

In Section~\ref{sec:2}, the equations of the usual single-phase models for ladle stirring are introduced. The definitions of the four gas fraction formulas, as well as some modifications of the 
modeling assumptions, are also discussed. Section~\ref{sec:3} describes the results of the numerical simulations for the 2d axial-symmetrical configuration.
The 3d application on the laboratory-scale ladle stirring with two eccentric gas nozzles (\cite{Palovaara2018}) is presented in Section~\ref{sec:4}.
Finally, Section~\ref{sec:5} presents the conclusions.

\section{Theoretical and numerical considerations}\label{sec:2}

\subsection{Definition of the model}\label{sec:21}

Cylindrical ladles with one or two nozzles in the bottom are considered in this paper. 
In Section~\ref{sec:3}, a 2d axial-symmetrical configuration 
and in Section~\ref{sec:4}, a 3d setup are studied. For the 
sake of brevity, the descriptions presented in this section are only 
for the 2d axial-symmetrical case. They apply similarly for the 3d 
situation, see also Section~\ref{sec:4} for some necessary adaptations. 

\begin{figure}[t!]
	\begin{center}
		\includegraphics[width=0.3\textwidth]{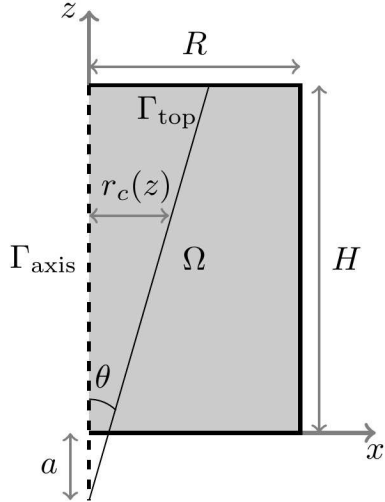}
	\end{center}
	\caption{Axial-symmetrical fluid domain (for Section~\ref{sec:3}).} \label{fig:modeldomain}
\end{figure}

\paragraph{Geometry and notations} The flow is assumed to be axial-symmetrical and the cylindrical coordinates are used. Let $\Omega  \subset \mathbb{R}^2$ designates the 2d fluid domain in a vertical plane of the cylinder passing through the central nozzle. The top boundary $\Gamma_{\text{top}}$ corresponds to the free surface of the fluid, and the vertical left wall is called $\Gamma_{\text{axis}}$. The geometry and notations are given in Figure~\ref{fig:modeldomain}.

Let the fluid velocity $\bu$ and pressure $p$ be time-dependent functions in the domain $\Omega$. The time interval on which the flow is computed is denoted by $(0,T_{\text{end}}]$. The liquid and gas density are $\rho_l$ and  $\rho_g$, with $\rho_l \gg \rho_g$, and the liquid viscosity is $\mu$. Finally, the stress and velocity deformation tensors are given by $\mathbb{S} = 2\mu\mathbb{D}(\bu) + p \mathbb{I} $ with  $\mathbb{D} = \frac{\nabla \bu  + (\nabla \bu)^T}{2} $.

\paragraph{PDEs and usual assumptions}The single-phase models for ladle stirring are based on the incompressible Navier--Stokes equations:
\begin{align}
\rho \parci{\bu}{t} + \rho ( \bu \cdot \nabla) \bu -  & 2\nabla \cdot ( \mu \mathbb{D} (\bu)) + \nabla p  =  \boldsymbol{f_v}      \label{eq:nse_ladle}\\
\nabla \cdot \bu  &= 0.    \nonumber
\end{align}
The volume force is, in most studies, equal to:
\begin{align}
	\boldsymbol{f_v}  = \rho_l g \alpha\  \boldsymbol{e_z},		\label{eq:volume_force}
\end{align}
except in~\cite{Grevet1982, Sahai1982,Woo90,Mukhopadhyay2001,Goldschmit2001}, where it is $\rho g \alpha$ or $\rho g$. 

In addition, an inhomogeneous density $\rho$ is defined. It is equal to the gas-liquid mixture density inside the gas plume, and to the liquid density outside the plume:
\begin{align}
\rho =
\left\{
\begin{array}{ll}
 \rho_g \alpha + \rho_l (1-\alpha) & \text{if}  \  r\leq r_c(z),  \\
 \rho_l      											  & \text{if} \  r\geq r_c (z),   
\end{array}  \right.  \label{eq:rho_plume}
\end{align}
where $r_c(z) = \tan (\theta) \left(  z+a \right)$ is the radius of the plume at height $z$, $\theta$ and $a$ being the apex and origin of the conical plume, respectively (Figure~\ref{fig:modeldomain}). The different gas fraction formulas $\alpha$ are discussed 
in Section~\ref{sec:22}.

\paragraph{Boundary conditions} Several types of boundary conditions are usually applied. In practice, the fluid velocity at the walls and the bottom part of the ladle is zero, so that homogeneous Dirichlet conditions are used:
\begin{align}
 \bu &= \boldsymbol{0}    \quad   \text{on}  \  \partial\Omega \setminus \{ \Gamma_{\text{axis}} \cup \Gamma_{\text{top}} \}.  \label{eq:bc_homogenousD2D}
\end{align}
The liquid bath surface is a free surface, normally covered by a slag layer, and subject to an unsteady movement, whose intensity depends on the gas flow rate. In the quasi-single phase models, the slag is not modeled explicitly. Instead, a free slip condition with no penetration is applied:
\begin{align}
\bu \cdot \boldsymbol{n}  &= 0   &  &\text{on}  \    \Gamma_{\text{top}}, \label{eq:bc_slip} \\
\boldsymbol{n}^T   \mathbb{S}  \ \boldsymbol{t}  &= 0   &  &\text{on}  \    \Gamma_{\text{top}},  \nonumber
\end{align}
where $\boldsymbol{n}$ and $\boldsymbol{t}$ are the unit normal and tangential vectors at the boundary.
This reduces unphysical flow braking close to the top surface, which would have been induced by homogeneous Dirichlet conditions.
On the vertical left axis, the axial symmetry naturally imposes the following conditions:
\begin{align*}
 \bu \cdot \boldsymbol{n}  &= 0   &  &\text{on}  \    \Gamma_{\text{axis}}, \\
\boldsymbol{n}^T   \mathbb{S}  \ \boldsymbol{t}  &= 0   &  &\text{on}  \    \Gamma_{\text{axis}}.  \nonumber
\end{align*}

\paragraph{Initial conditions.} At $t=0$, the fluid is assumed to be at rest in $\Omega$, i.e., 
the initial velocity is zero:
\begin{align*}
\bu(0,\bx) &= \boldsymbol{0}   &  &\text{in}  \    \Omega.  
\nonumber
\end{align*}

\subsection{Gas fraction $\alpha$}\label{sec:22}

\paragraph{Definitions} As mentioned in  Section~\ref{sec:1}, the
gas fraction formulas from~\cite{Sahai1982,Balaji1991,Woo90,Mukhopadhyay2001} have been proved to be superior to the 
formulas from~\cite{Debroy1978,Grevet1982}. However, to the best of the authors' knowledge, a comparison of the former definitions is still needed, in order to help choosing the most appropriate formula for practical applications.

In the following, the notation $\alpha$ refers to the gas fraction, regardless of the formula applied, while indices refer to specific definitions from the literature.

Sahai and Guthrie~\cite{Sahai1982} defined $\alpha$ as a constant:
\begin{align}
\alpha_1 = \frac{Q }{\pi r_{\text{av}}^2 U_P}, \label{eq:alpha_sahai}
\end{align}
where $U_P$ is the plume velocity, and $r_{\text{av}}$ is the average of the plume radius. The present work retains the plume velocity from Mazumdar et al.~\cite{Mazumdar1993}, $ U_P = 4.4 \frac{Q^{1/3}H^{1/4}}{R^{1/4}} $, which is an improved version of the original paper. The average radius can be computed as $r_{\text{av}}=  \frac{1}{2} \left(  \tan (\theta) \left( 2a+H    \right)  \right) $. As pointed out by the authors, this definition corresponds to the average gas fraction in the whole plume.

In~\cite{Balaji1991}, the gas fraction is defined as:
\begin{align}
\alpha_2 = \frac{Q - \pi r_c^2(z) \alpha_2 (1-\alpha_2) U_S}{\pi r_c^2(z)U_P}, \label{eq:alpha_maz}
\end{align}
where $U_S$ is the slip velocity between gas bubbles and the liquid. Its value ranges between $0.1~\unitfrac{m}s$ (\cite{Mazumdar1995b}) and $0.6~\unitfrac{m}s$ (\cite{Mazumdar1992}). In this work, an intermediary value of 
$0.4~\unitfrac{m}s$ is applied,~\cite{Grevet1982,Woo90}. In Equation~\eqref{eq:alpha_maz}, $\alpha_2$ depends on the vertical coordinate $z$ through $r_c$. The additional term in the top part of the fraction is derived from the so-called drift-flux model,~\cite{Grevet1982}.

It is possible to solve Equation~\eqref{eq:alpha_maz} analytically, as a solution of a 2nd-order polynomial:
\begin{align}
\alpha_2 = \frac{1}{2} \left( \left( \frac{U_P}{U_S} +1  \right) -  \sqrt{ \left(  \frac{U_P}{U_S} +1\right)^2 -\frac{4Q}{\pi r_c^2(z)U_S} }  \right).  \nonumber 
\end{align}

Note that this is valid only for $ z \geq z_C$, where $z_C = \frac{1}{\tan(\theta)}  \left( \sqrt{\frac{4Q}{\pi U_S (\frac{U_P}{U_S} +1)^2} }   \right)  - a $. One can however extend the definition to the small heights, such that:
\begin{align}
\alpha_2 =
\left\{
\begin{array}{ll}
 \frac{1}{2}  \left( \frac{U_P}{U_S} +1  \right)  & \text{if}  \  z\leq z_C,  \\
 \frac{1}{2} \left( \left( \frac{U_P}{U_S} +1  \right) -  \sqrt{ \left(  \frac{U_P}{U_S} +1\right)^2 -\frac{4Q}{\pi r_c^2(z)U_S} }  \right)      											  & \text{if} \  z\geq z_C,   
\end{array}  \right.  \label{eq:alpha_maz2}
\end{align}
To the best of our knowledge, the analytical solution~\eqref{eq:alpha_maz2} and the lack of definition of~\eqref{eq:alpha_maz} in small heights
were not addressed in the literature so far.

Finally, the formulas applied in~\cite{Woo90,Mukhopadhyay2001} originally come from the experimental work of Castillejos and Brimacombe~\cite{Castillejos1987,Castillejos1989}. By assuming that the gas fraction follows a Gaussian distribution and using experimental correlations, the latter derived two variants $\alpha_3$ and $\alpha_4$ of the form:
	\begin{align}
	\alpha(r,z) =
	\left\{
	\begin{array}{ll}
	\ c_1  z^{\beta} \exp \left[  -0.7 \left( \frac{r}{c_2 z^{\delta}} \right)^{2.4}    \right] &\  \text{if} \ z < z_0, \\[2ex]
	\ c_3  z^{\gamma} \exp \left[  -0.7 \left( \frac{r}{c_2  z^{\delta}} \right)^{2.4}    \right] &\  \text{if} \ z \geqslant z_0, \\
	\end{array}
	\right. 		  \label{eq:alpha_brima}
	\end{align} 
where the constants depend on the gas flow rate, the nozzle diameter, and the densities of gas and liquid. More details can be found in~\cite{Castillejos1987,Castillejos1989}. Here, unlike Equations~\eqref{eq:alpha_sahai} and~\eqref{eq:alpha_maz2}, the gas fraction depends
on both $r$ and $z$, and it is discontinuous in the height.

Before computing the solution of the model, the four formulas are compared in order to identify the differences between them. The parameters for the numerical application are listed in Table~\ref{table:parameters} and the corresponding values for $\alpha_3$ and $\alpha_4$ are given in Table~\ref{table:constants_brima}.

\begin{table}[t!]
	\begin{center}
		\caption{Parameters of the ladle stirring model (\cite{Woo90}).}
		\begin{tabular}{c |c|c|c|c}
					\hline
			$H$ &	$R$ &$\theta$ 	&$a$ 	& 	$U_S$ 	 	\\ 
			0.6 $\unit{m}$ &   0.3 $\unit{m}$  &  10 &  0.08 $\unit{m}$&    0.4 $\unit{m/s}$   \\ \hline \hline
			 $\rho_l$ & $\rho_g$ &  $d $ & $r_{\text{av}}$  &       $Q$ 		\\
			  1000 $ \unit{kg/m^3}$ &     1  $\unit{kg/m^3}$  &    12.7 $\unit{mm}$ & 0.067  $\unit{m}$  &    13 $\unit{l/min} $ \\ \hline
		\end{tabular}
		\label{table:parameters}
	\end{center}
\end{table}

\begin{table}[t!]
	\caption{Constants of Equation~\eqref{eq:alpha_brima} for $\alpha_3$ and $\alpha_4$ (\cite{Castillejos1987,Castillejos1989}).}
	\begin{center}
		\begin{tabular}{c |c c c c c c c}
			& $c_1 $ &$c_2 $ &$c_3 $ &$ z_0$ &$\beta $ &$ \gamma  $ &$ \delta  $ \\ \hline $\alpha_3$  & 29.8785	& 0.0934	& 1.2114           & 0.016			& $ -0.218$      & $  -0.993$ & 0.48          \\
			$\alpha_4$    & 52.9798		 &  0.0781	 &	1.4405	  &   0.0141 & $-0.094	$	&  		$-0.94$		& 0.51			
		\end{tabular}
		\label{table:constants_brima}
	\end{center}
\end{table}

\begin{figure*}
	\begin{center}
	\includegraphics[width=0.9\textwidth]{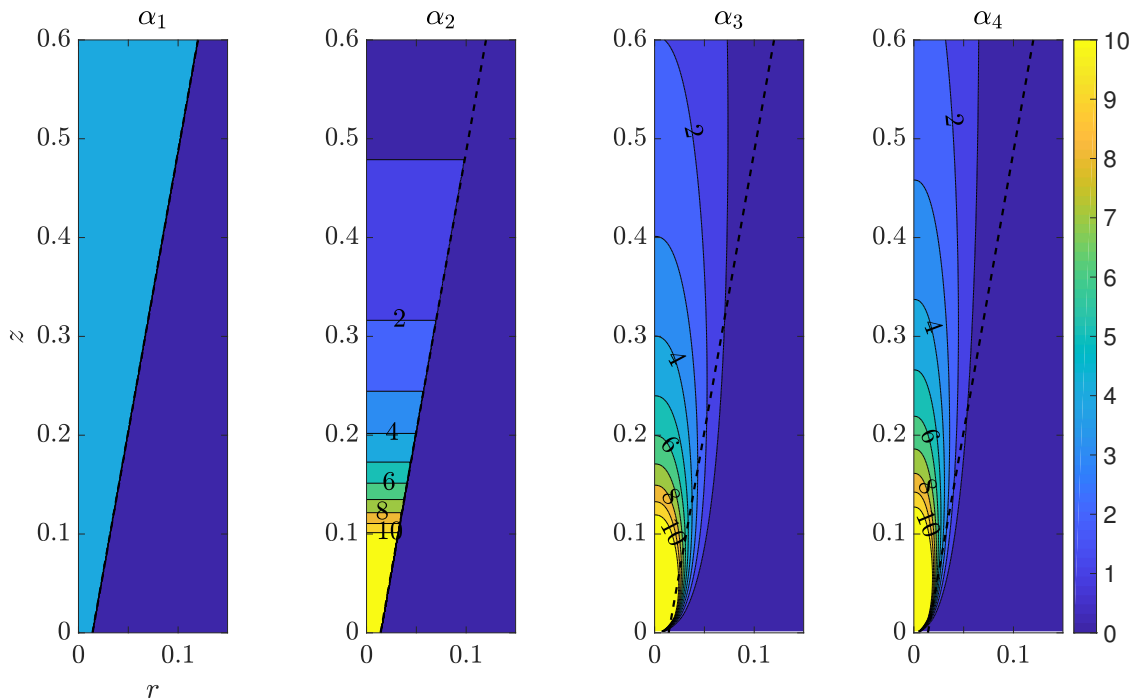}
	\end{center}
	\caption{Isolines of the four different formulas for $\alpha$.} \label{fig:alpha_isolines}
\end{figure*}

Figure~\ref{fig:alpha_isolines} illustrates the isolines of the gas fraction field according to each definition.
One can first notice that, in $\alpha_3$ and $\alpha_4$, the plume shape is naturally described by the exponential, whereas in $\alpha_1$ and $\alpha_2$, the cone $r_c(z)$ is needed to restrict the gas fraction to the plume region. However, the cone gives a sharp shape to the plume and creates a discontinuity which may be not representative of the physical reality. 
In the case of $\alpha_3$ and $\alpha_4$, the shape of the plume also increases with higher gas flow rate $Q$, which is closer to the physical reality, than a conical plume independent of $Q$.
It is interesting to note however that the difference between the strict conical plume and the one obtained with $\alpha_3$ and $\alpha_4$ is relatively small. 
In most of the plume region, $\alpha$ is less than 10\% with all formulas. While $\alpha_1$ is constant in the plume ($\sim4.8\%$) and $\alpha_2$ is stratified in the height, the formulas $\alpha_3$ and $\alpha_4$ vary smoothly in both directions $r$ and $z$, which corresponds maybe better to reality. Note that there is not a big difference between $\alpha_3$ and $\alpha_4$, although the second one yields slightly higher values and a narrower plume. 

Close to the nozzle, the gas fraction increases rapidly to more than 10\% in the three cases $\alpha_2$, $\alpha_3$, and $\alpha_4$. While $\alpha_2$ is fixed to a constant in this region (Eq.~\eqref{eq:alpha_maz2}), $\alpha_3$ and $\alpha_4$ blow up at $(0,0)$, which is clearly not representative of the physical reality.

This preliminary discussion shows that the order of magnitude and the shape of the different gas fraction fields are not fundamentally different and that both formulas~\cite{Castillejos1987,Castillejos1989} are very similar. In the rest of this study, only $\alpha_1$, $\alpha_2$, and $\alpha_3$ are considered.

\subsection{Modified modeling assumptions}\label{sec:23}

In this paper, several modifications are proposed to the existing models. 
	First, the variable density \eqref{eq:rho_plume} is simplified to:
	\begin{align*}
	\rho = \rho_l 
	\end{align*} 
in the whole domain, such that standard numerical solvers for the incompressible Navier--Stokes equations can be used. The distinction between the plume region and the rest of the liquid is modeled with the gas fractions, leading to the new definitions:
\begin{align}
\alpha'_1=
\left\{
\begin{array}{ll}
\alpha_1 & \text{if}  \  r\leq r_c(z),  \\
0  											  & \text{if} \  r > r_c (z),   
\end{array}  \right.  \label{eq:sahai2}
\end{align}
and
\begin{align}
\alpha'_2=
\left\{
\begin{array}{ll}
\alpha_2 & \text{if}  \  r\leq r_c(z),  \\
0  											  & \text{if} \  r > r_c (z),   
\end{array}  \right.  \label{eq:maz3}
\end{align}
respectively. In order to simplify the notations, Equations~\eqref{eq:sahai2} and~\eqref{eq:maz3} will be referred to as $\alpha_1$ and $\alpha_2$.

The formula~\eqref{eq:alpha_brima} remains unchanged for $\alpha_{3}$. Finally, the volume force~\eqref{eq:volume_force} is replaced by:
\begin{align}
	\boldsymbol{f_v}  =  ( - \rho_l g +  \rho_l g \alpha)   \boldsymbol{e_z},	 \label{eq:volume_force2}
\end{align}
where $\alpha$ is either equal to $\alpha_1$, $\alpha_2$, or $\alpha_3$. This volume force accounts for the gravity on the liquid in addition to the buoyancy force generated by the gas plume. This allows to capture the hydrostatic pressure and gives more realistic values  for $p$ than Eq.~\eqref{eq:volume_force}, even if it does not change the flow pattern.

\subsection{The $k$-$\epsilon$ turbulence model}

Ladle stirring is known to be a turbulent flow (\cite{Mazumdar1995}). The main approach 
applied in literature on ladle stirring to resolve the turbulence is the standard $k$-$\epsilon$ model:
\begin{align*}
\mu_t &= \rho C_{\mu} \frac{k^2}{\epsilon}, \\
\rho \parci{k}{t} + \rho (\bu \cdot \nabla) k &= \nabla \cdot [(\mu + \frac{\mu_t}{\sigma_k} ) \nabla k  ] + G - \rho \epsilon,\\
\rho \parci{\epsilon}{t} + \rho (\bu \cdot \nabla) \epsilon &= \nabla \cdot [(\mu + \frac{\mu_t}{\sigma_{\epsilon}} ) \nabla \epsilon  ] + C_{1}\frac{\epsilon}{k}G - C_{2}\rho\frac{\epsilon ^2}{k},  \\
G& = \mu_t [\nabla \bu \colon  (\nabla \bu + (\nabla \bu)^T  )] .
\end{align*}
The turbulent viscosity $\mu_t$ is then added to the liquid viscosity $\mu$ in the viscous term of the Navier--Stokes equations~\eqref{eq:nse_ladle}. Default values are used for the constants of the $k$-$\epsilon$ model~\cite{Woo90}.

\begin{table}[t!]
	\begin{center}
		\caption{Axial-symmetrical configuration: Size and number of cells used in the different meshes in the convergence study.}
		\begin{tabular}{c|c|c|c}
			Mesh number & 0 & 1 & 2 \\
			\hline
			Mesh size  ($\unit{m}$) & 0.012 & 0.006 & 0.003 \\
			Grid & 25$\times$50 & 50$\times$100 & 100$\times$200 \\
			Number of cells & 2500 & 5000 & 20000 \\
			Deg. of freedom & 11628 & 45753 & 181503 \\ \hline
		\end{tabular}
		\label{table:convstudy2}
	\end{center}
\end{table}

\section{The axial-symmetrical configuration}\label{sec:3}

\subsection{Numerical procedure}

The models were numerically solved with the Finite Element Method (FEM) using the commercial software {\sc Comsol Multiphysics\textsuperscript{\textcopyright}} (version 5.3a). The rectangular ladle (Figure~\ref{fig:modeldomain}) was meshed with quadrilaterals.
Three meshes were applied to check the convergence of the numerical solutions. The Taylor--Hood pair $Q_2/Q_1$ of finite elements was used for the velocity and the pressure, i.e., the velocity is approximated 
with continuous piecewise biquadratic functions and the pressure with 
continuous piecewise bilinear functions. This pair of inf-sup stable spaces is one of 
the most popular ones, \cite{Joh16}. 
The mesh size, number of 
mesh cells and degrees of freedom of the velocity and pressure fields are given in Table~\ref{table:convstudy2}.
\begin{figure*}
	\begin{center}
		\includegraphics[scale=0.3]{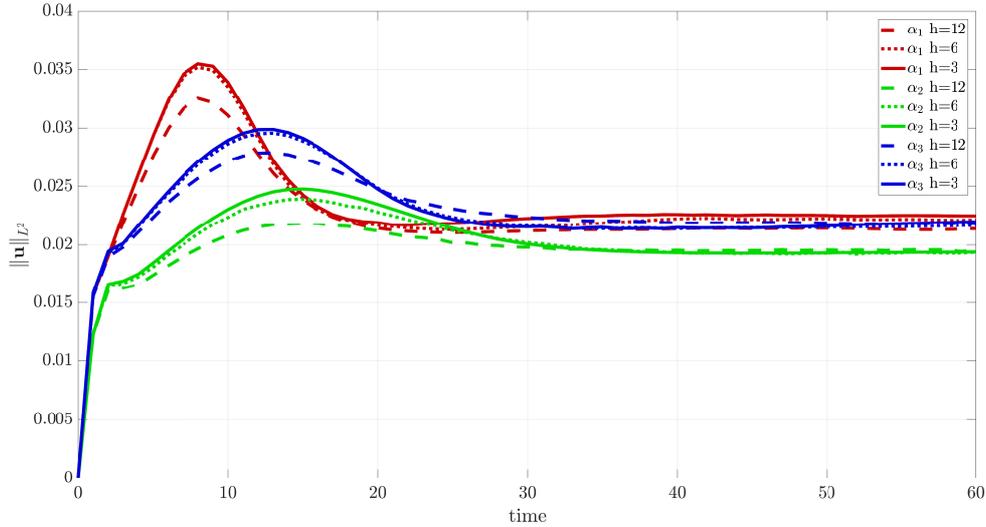}
	\end{center}
	\caption{Axial-symmetrical configuration: $L^2$-norm of the velocity with the different gas fractions and meshes.} \label{fig:results_norm} 
\end{figure*}

No stabilization of the convection term in the Navier--Stokes equations is applied, since additional viscosity is added through the turbulent viscosity. 
The time discretization scheme is BDF2. The time-stepping is adaptive, with a maximum time step of $2~\unit{s}$. Finally, $T_{\text{end}}$ is set to $60~\unit{s}$.

\subsection{Results and discussions}

Mesh convergence and solution stationarity were verified using the $L^2$-norm of the velocity field:
\begin{align*}
	\| \bu \|_{L^2} = \left(    \int_{\Omega}  \| \bu\|^2\ d\boldsymbol x  \right)^{1/2}.
\end{align*}
The results are given in Figure~\ref{fig:results_norm}. For all three gas fractions, the steady-state is considered to be reached between $30~\unit{s}$ and $40~\unit{s}$. It can also be seen that the values obtained with the different mesh sizes are relatively close, especially in the steady-state. An intermediary mesh size of $6~\unit{mm}$ can thus be considered to be sufficient for future computations.

\begin{figure*}
	\begin{center}
		\includegraphics[scale=0.6]{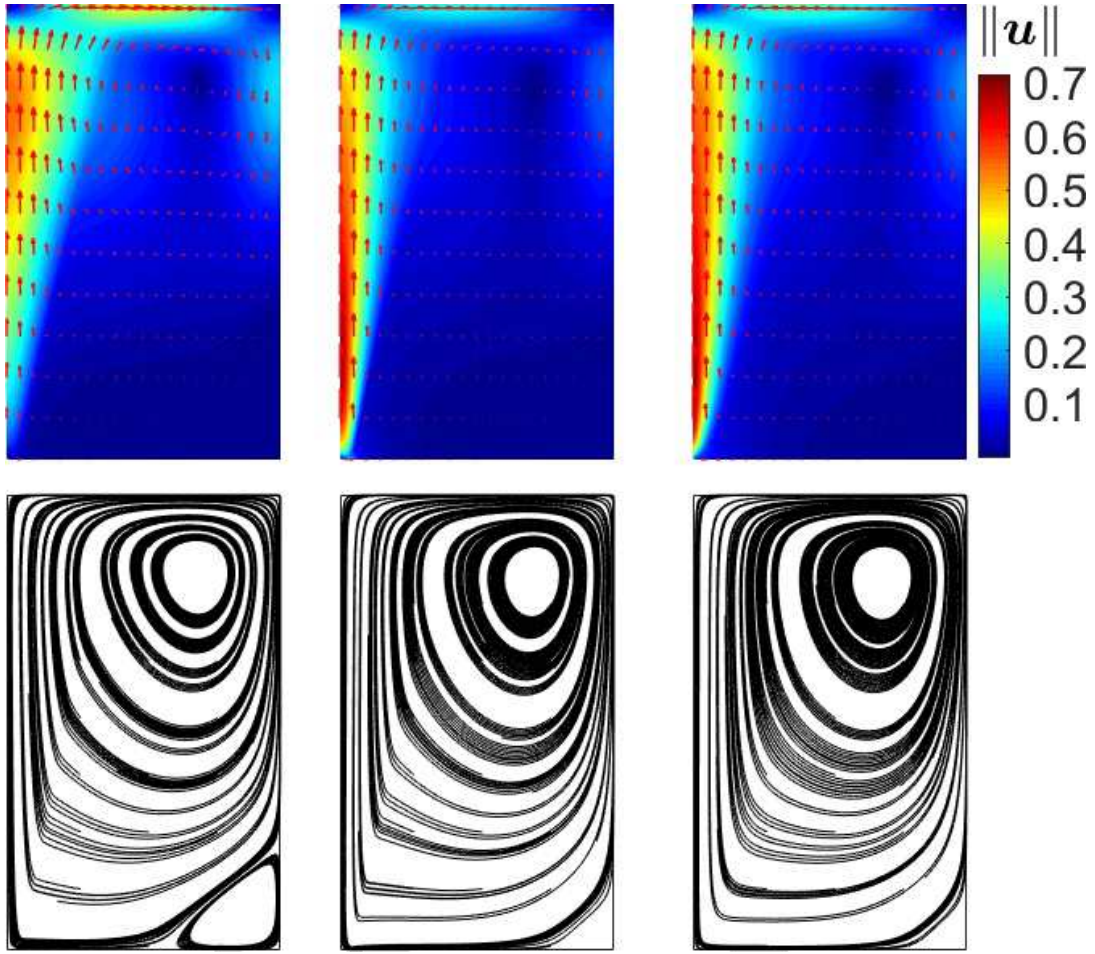}
	\end{center}
	\caption{Axial-symmetrical configuration: Velocity fields and streamlines. Left to right: $\alpha_1$, $\alpha_2$, and $\alpha_3$.} \label{fig:results_fields} 
\end{figure*}

The computed velocity fields on the finest meshes are displayed in Figure~\ref{fig:results_fields}. The velocity field reveals the effect of the gas through the volume force~\eqref{eq:volume_force2}. A strong upward flow is generated close to the left boundary. Its intensity close to the nozzle is higher with $\alpha_2$ and $\alpha_3$ than with $\alpha_1$, which can clearly be assigned to the higher gas fraction in this zone (Figure~\ref{fig:alpha_isolines}). On the contrary, on the top left side of the domain, the velocity is slightly higher with $\alpha_1$ than with $\alpha_2$ and $\alpha_3$. In a similar way, this is due to the higher gas fraction for $\alpha_1$ ($\sim4.8\%$) in comparison with the two others ($< 2\%$).
Far from the left boundary, the velocity fields obtained with the three gas fractions are relatively similar, in the sense that they all produce one vortex located in the upper right region. 

\begin{figure*}
	\begin{center}
		\includegraphics[width=0.8\textwidth]{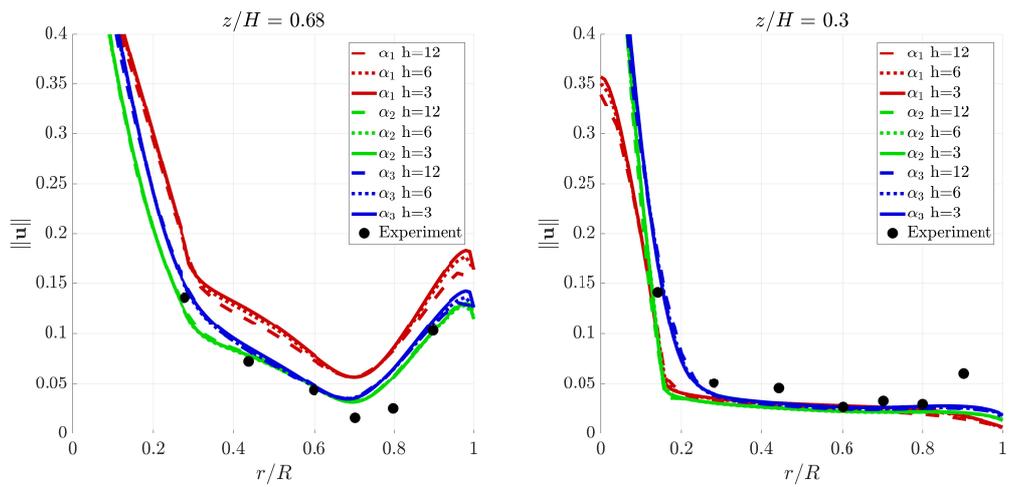}
	\end{center}
	\caption{Axial-symmetrical configuration: Comparison of the Euclidean norm of the velocity at two different heights of the bath.} \label{fig:results_comparison} 
\end{figure*}

A comparison of the velocity magnitude with experimental measurements from~\cite{Woo90} at two different heights is given in Figure~\ref{fig:results_comparison}. Similar velocity profiles are observed for all proposals of the gas fraction, but with slight differences in the amplitude. The gas fractions $\alpha_2$ and $\alpha_3$ match better with the experimental measurements, than $\alpha_1$, especially in the upper region of the domain ($z/H=0.68$). 

In~\cite{Woo90}, additional measurements of the velocity and the turbulent kinetic energy are available.  However, they are not reproduced here, because a higher discrepancy is observed between the present simulations and the reported experimental measurements, especially at the level $z/H=0.98$. Indeed, close to the boundaries, the $k$-$\epsilon$ turbulence model employs wall functions, which are differently implemented in~\cite{Woo90} and in the software used here.
Thus, a meaningful comparison of the results close to the boundary is not possible.

\section{Application to a real laboratory-scale ladle in 3d}\label{sec:4}

In this application, the geometry corresponds to a real laboratory-scale water ladle with two eccentric nozzles,~\cite{Palovaara2018}. The notations are similar to the ones introduced previously, except that, in this case, a Cartesian space frame is used because the axial-symmetrical assumption does not hold. Table~\ref{table:parameters3D} lists the parameters for this application.

\begin{table}[t!]
	\begin{center}
		\caption{Laboratory-scale ladle in 3d: Parameters of the ladle stirring model.}
		\begin{tabular}{c c c c c c c c}
			\hline
			$H$ 			&  $R_{\text{top}}$ 			&  $R_{\text{bot}}$ 	&$Q$ &$x_{n1}$ 	&$y_{n1}$ 	& $x_{n2}$ 	& $y_{n2}$	\\ \hline
			0.65 $\unit{m}$ &   0.29 $\unit{m}$ &    0.27 $\unit{m}$ &	17 $\unit{l/min} $ &
			-0.105 $\unit{m}$ &	-0.105 $\unit{m}$ &  -0.105 $\unit{m}$&  0.105 $\unit{m}$ \\ \hline
		\end{tabular}
		\label{table:parameters3D}
	\end{center}
\end{table}

\begin{figure}[t!]
	\begin{center}
		\includegraphics[width=.6\textwidth]{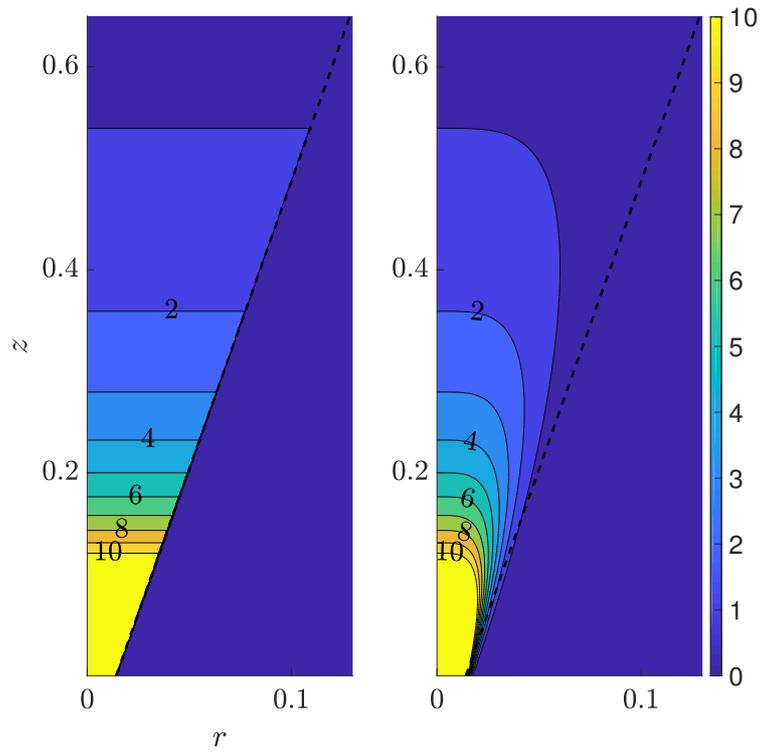}
	\end{center}
	\caption{Isolines of the gas fraction in a vertical plane passing through one of the nozzles. Left: non-smoothed formula~\eqref{eq:alpha_maz3}. Right: smoothed formula~\eqref{eq:alpha_maz4}.} \label{fig:alpha_isolines2}
\end{figure}

It should be noted that the nozzles in this model are fictive: they correspond to the origin of the plume cones. The volume force in this case is defined as:
\begin{align*}
\boldsymbol{f_v}  = \left(    -\rho g + (\alpha_{n1}+\alpha_{n2})\rho g \right)\  \boldsymbol{e_z},		
\end{align*}
where $\alpha_{ni}$, $i=1,2$, corresponds to the gas fraction in the plumes generated by nozzle $i$.
Given the previous results, the formula $\alpha_2$ from Equation~\eqref{eq:alpha_maz2} seems to be an appropriate choice for the 3d application. However, this formula needs to be adapted to the Cartesian coordinates.
The gas fraction of each nozzle $\alpha_{ni},\ i=1,2$, is then defined as:
	\begin{align}
	\alpha_{ni}(x,y,z)=
	\left\{
	\begin{array}{ll}
	\alpha_2(z) & \text{if}  \  (x-x_{ni})^2+(y-y_{ni})^2  \leqslant r_c^2(z),  \\
	0  											  & \text{else},   
	\end{array}  \right.  \label{eq:alpha_maz3}
	\end{align} 
where $(x_{ni},y_{ni})$ is the center of nozzle $i =1,2$. In order to make the transition between the gas plume and the liquid smoother, $\alpha_{ni}$ is modified to:
\begin{equation}
\alpha_{ni}(x,y,z)=	\alpha_2(z) \exp\left( - b\left(\frac{(x-x_{ni})^2 + (y-y_{ni})^2  }{r_c(z)^2}\right)^2  \right). 		  \label{eq:alpha_maz4}
\end{equation}
This definition is more realistic from the physical point of view, since it reduces the sharp discontinuity found in Equation~\eqref{eq:alpha_maz3}. For comparison, the isolines of the gas fraction fields~\eqref{eq:alpha_maz3} and~\eqref{eq:alpha_maz4}, in an equivalent axial-symmetrical 2d case, are illustrated in Figure~\ref{fig:alpha_isolines2}.
The smoothness of the transition between the gas plume and the liquid can be set with the parameter $b$. In this study, $b=2$ was used.

The boundary conditions are identical to Equations~\eqref{eq:bc_homogenousD2D} and~\eqref{eq:bc_slip}, except that there is no symmetry axis anymore.
Moreover, since the flow is enclosed, an additional condition for the pressure is needed to ensure uniqueness of the solution. A node of the top surface was chosen with a relative pressure fixed to be~$0$.

\begin{figure}[t!]
	\begin{center}
		\includegraphics[width=0.5\textwidth]{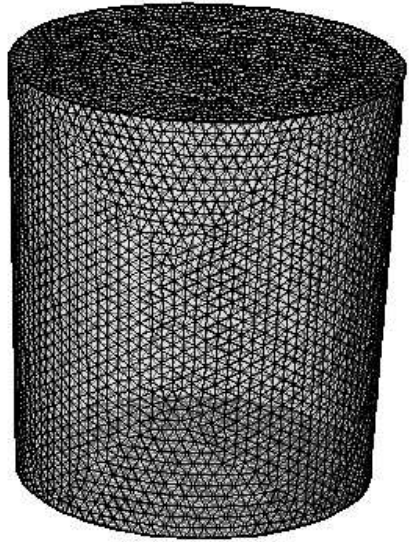} 
	\end{center}
	\caption{Laboratory-scale ladle in 3d: Mesh 2 ($h = 20~\unit{mm}$).}   \label{fig:3dtank_mesh}
\end{figure}

\begin{table}[t!]
	\begin{center}
		\caption{Laboratory-scale ladle in 3d: Mesh parameters.}
		\begin{tabular}{c| c|c|c}
			\hline
			\multicolumn{2}{c|}{Mesh number} &  1 & 2 \\ \hline
			\multicolumn{2}{c|}{Mesh size  ($\unit{m}$) } & 0.03 & 0.02 \\
			\multicolumn{2}{c|}{Number of elements } & 100 377 & 343 846\\ \hline
			\multirow{2}{*}{\shortstack{Deg. of\\ freedom}} &    Velocity      &  415 137     & 1 405 701\\ 
			&   Pressure	 &   17 857    &  59 742 \\ \hline
		\end{tabular}
		\label{table:convstudy}
	\end{center}
\end{table}

\begin{figure}[t!]
	\begin{center}
		\includegraphics[width=0.6\textwidth]{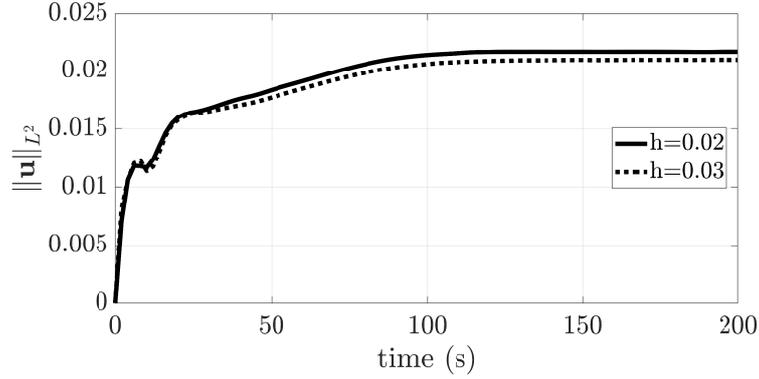}
	\end{center}
	\caption{Laboratory-scale ladle in 3d: $L^2$-norm of the velocity field.} \label{fig:results_norm2} 
\end{figure}

The fluid domain was meshed with unstructured tetrahedral cells, as illustrated in Figure~\ref{fig:3dtank_mesh}. More information about the meshes are given in Table~\ref{table:convstudy}. The Taylor--Hood pair $P_2/P_1$ of finite elements was used.
The default stabilization methods of {\sc Comsol Multiphysics\textsuperscript{\textcopyright}}
(streamline and cross diffusion) were applied for both the Navier--Stokes equations and turbulence equations. Finally, preliminary computations were performed to find an end time where a stationary solution is reached. 
The $L^2$-norms of the velocity fields are given in Figure~\ref{fig:results_norm2}, where it can be seen that the steady-state is reached shortly after $100~\unit{s}$. 
Consequently, $T_{\text{end}}$ is set to $200~\unit{s}$.

\begin{figure*}[t!]
	\begin{center}
		\includegraphics[width=0.98\textwidth]{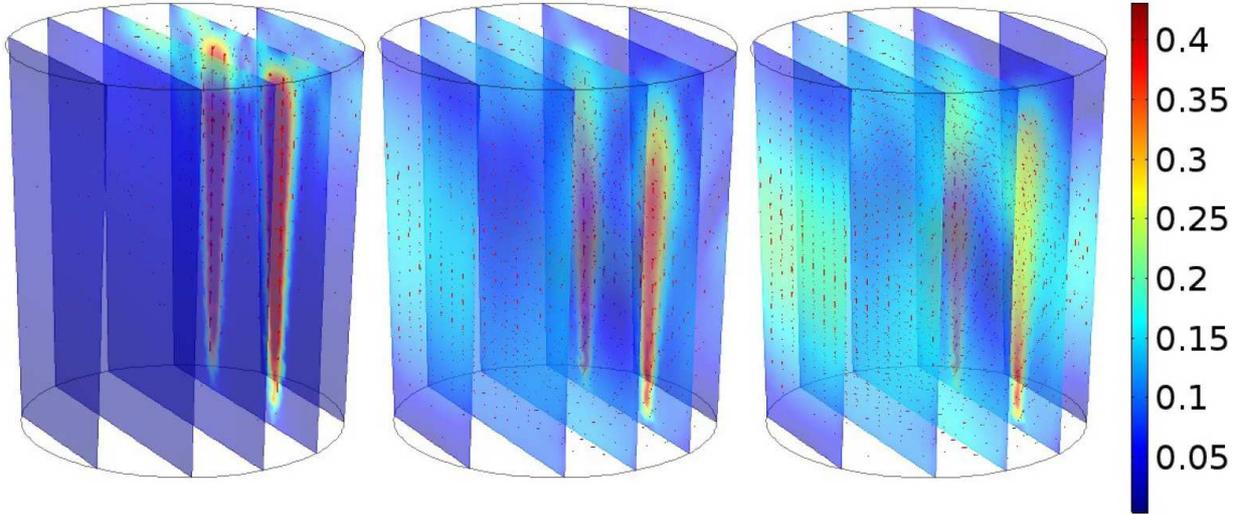} 
	\end{center}
	\caption{Laboratory-scale ladle in 3d: Velocity fields at $t = 2,25,200~\unit{s}$.} \label{fig:results_fields2}
\end{figure*}

The velocity field at $t=2,\ 25,$ and $200~\unit{s}$ is illustrated in Figure~\ref{fig:results_fields2}. As expected, the volume force~\eqref{eq:volume_force2} produces the desired gas plume effect: an upward flow is generated from the position of the 
nozzles at the bottom to the top surface. Its intensity decreases from the bottom, close to the nozzle, to the top, while its radius expands with $z$. This flow pattern is qualitatively similar to the ones reported in~\cite{Mazumdar1995}.

\begin{figure*}[t!]
	\begin{center}
		\includegraphics[width=0.6\textwidth]{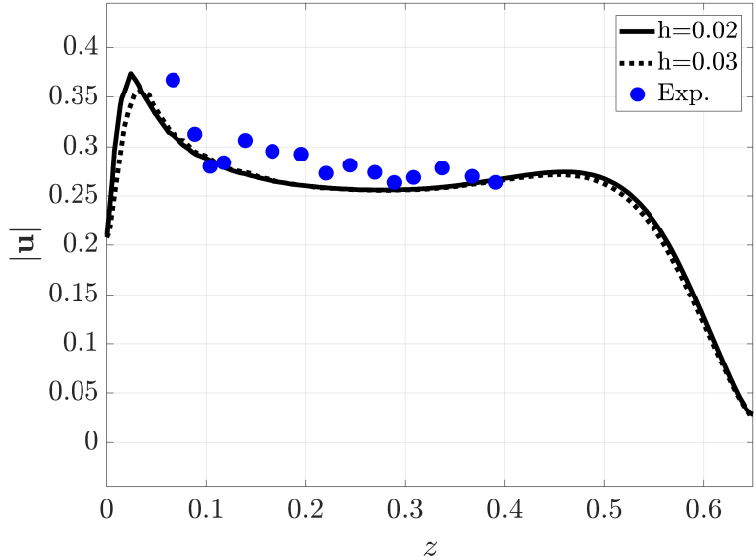} 
	\end{center}
	\caption{Laboratory-scale ladle in 3d: Euclidean norm of the velocity in the centerline of the gas plume generated by nozzle 1 (i.e., along the vertical line going from $(x_{n1},y_{n1},0) $ to $(x_{n1},y_{n1},0.65) $) compared with experimental measurements in~\cite{Palovaara2018}.} \label{fig:results_centerlinevelo} 
\end{figure*}

Figure~\ref{fig:results_centerlinevelo} compares the computed velocity magnitude at the central line of one of the gas plumes from bottom to top, with experimental measurements conducted in~\cite{Palovaara2018}. 
The velocity profile is similar to the observations reported in \cite{Mazumdar1995}: starting at a high value close to the nozzle (jet zone dominated by the kinetic energy of the gas), it slowly decreases a few decimeters above the nozzle and remains constant in most of the bath height (plume zone dominated by the buoyancy energy). The decrease close to the surface is due to the boundary condition. At this level, a free surface modeling would have been more appropriate, but this needs a more complex and computationally expensive approach, such as moving meshes or multiphase flows. 
All in all, the velocity computed at the center of the gas plume is in reasonable agreement with experimental measurements.

\section{Conclusions}\label{sec:5}

In this study, a modified single-phase model for ladle stirring has been derived from existing `quasi-single phase' models from literature. Some changes in the modeling assumptions have been introduced to allow a more straightforward implementation of the model in standard incompressible flow solvers. Furthermore, the shapes of the gas fraction were clarified thanks to, on the one hand, the analytical solution~\eqref{eq:alpha_maz2}, and, on the other hand, a comparison of the isolines of the gas fraction. 

Although the comparison shows that their order of magnitude and their shape are quite similar, some
differences are still observable: the formula $\alpha_1$ (\cite{Sahai1982}) produces higher gas fractions in the top region of the ladle, than $\alpha_2$ and $\alpha_3$ (\cite{Balaji1991,Castillejos1987}). 
This is reflected in the numerical results: in the first case, the computed velocity magnitude is higher than the experimental measurements, while the two other formulas give reasonable agreement with measurements available in literature. They can thus be equivalently used in practical applications.

Subsequently, the gas fraction $\alpha_2$ has been applied for a real laboratory-scale 3d ladle with two eccentric nozzles. The simulation results show that the present model is appropriate from both qualitative and quantitative perspectives. Indeed, the velocity profile in the bath corresponds with results reported in the literature and the velocity at the centerline of the plume is in a relative good agreement with recent experimental measurements~\cite{Palovaara2018}.

In summary, we conclude that the accuracy of the results computed with the present single-phase model using $\alpha_2$ from \eqref{eq:maz3} or $\alpha_3$ form \eqref{eq:alpha_brima} is 
sufficient for employing it as an efficient
model in simulations of optimal flow control problems.

\section*{Acknowledgments}
Funding: This work was supported by the European Union's Horizon 2020 research and innovation programme [Marie Sk\l{}odowska-Curie grant agreement No. 675715 (MIMESIS)].


\section*{References}
\begin{thebibliography}{10}
	\expandafter\ifx\csname url\endcsname\relax
	\def\url#1{\texttt{#1}}\fi
	\expandafter\ifx\csname urlprefix\endcsname\relax\def\urlprefix{URL }\fi
	\expandafter\ifx\csname href\endcsname\relax
	\def\href#1#2{#2} \def\path#1{#1}\fi
	
	\bibitem{Sir87}
	L.~Sirovich, Turbulence and the dynamics of coherent structures. {I}.
	{C}oherent structures, Quart. Appl. Math. 45~(3) (1987) 561--571.
	
	\bibitem{NMT11}
	B.~R. Noack, M.~Morzynski, G.~Tadmor, Reduced-Order Modelling for Flow Control,
	Vol. 528, Springer Verlag, 2011.
	
	\bibitem{CIJS14}
	A.~Caiazzo, T.~Iliescu, V.~John, S.~Schyschlowa, A numerical investigation of
	velocity-pressure reduced order models for incompressible flows, J. Comput.
	Phys. 259 (2014) 598--616.
	
	\bibitem{Mazumdar1995}
	D.~Mazumdar, R.~I.~L. Guthrie, The physical and mathematical modeling of gas
	stirred ladle systems, ISIJ Intern. 35~(1) (1995) 1--20.
	
	\bibitem{Debroy1978}
	R.~Debroy, A.~K. Majumdar, D.~B. Spalding, Numerical prediction of
	recirculation flows with free convection encountered in gas-agitated
	reactors, App. Math. Model. 2~(3) (1978) 146--150.
	
	\bibitem{Grevet1982}
	J.~H. Grevet, J.~Szekely, N.~El-Kaddah, {An experimental and theoretical study
		of gas bubble driven circulation systems}, Intern. J. of Heat and Mass
	Transf. 25~(4) (1982) 487--497.
	
	\bibitem{Sahai1982}
	Y.~Sahai, R.~I.~L. Guthrie, {Hydrodynamics of gas stirred melts: Part II.
		Axisymmetric flows}, Metall. Trans. 13~(2B) (1982) 203--211.
	
	\bibitem{Balaji1991}
	D.~Balaji, D.~Mazumdar, {Numerical computation of flow phenomena in gas-stirred
		ladle svstems}, Steel Res. 62~(1) (1991) 16--23.
	
	\bibitem{Mazumdar2009}
	D.~Mazumdar, J.~W. Evans, Modeling of steelmaking processes, CRC Press, 2009.
	
	\bibitem{Woo90}
	J.~S. Woo, J.~Szekely, A.~H. Castillejos, J.~K. Brimacombe, A study on the
	mathematical modeling of turbulent recirculating flows in gas-stirred ladles,
	Metall. Trans. 21~(21B) (1990) 269--277.
	
	\bibitem{Bernard2000}
	R.~Bernard., R.~S. Maier, H.~T. Falvey, {A simple computational model for
		bubble plumes}, App. Math. Model. 24~(3) (2000) 215--233.
	
	\bibitem{Mazumdar1992}
	D.~Mazumdar, T.~Narayan, P.~Bansal, {Mathematical modelling of mass transfer
		rates between solid and liquid in high-temperature gas-stirred melts}, App.
	Math. Model. 16~(5) (1992) 255--262.
	
	\bibitem{Ganguly2004}
	S.~Ganguly, S.~Chakraborty, {Numerical investigation on role of bottom gas
		stirring in controlling thermal stratification in steel ladles}, ISIJ Intern.
	44~(3) (2004) 537--546.
	
	\bibitem{Mukhopadhyay2001}
	A.~Mukhopadhyay, P.~Deb, A.~Ghosh, B.~Basu, R.~Dutta, P.~Kumar, {Prediction of
		temperature in secondary steelmaking : mathematical modelling of fluid flow
		and heat transfer in gas purged ladle}, Steel Res. 72~(5) (2001) 192--199.
	
	\bibitem{Zhu96}
	M.-Y. Zhu, I.~Sawada, N.~Yamasaki, T.-C. Hsiao, Numerical simulation of
	three-dimensional fluid flow and mixing process in gas-stirred ladles, ISIJ
	Inter. 36~(5) (1996) 503--511.
	
	\bibitem{Goldschmit2001}
	M.~Goldschmit, A.~C. Owen, {Numerical modelling of gas stirred ladles},
	Ironmak. and Steelmak. 28~(4) (2001) 337--340.
	
	\bibitem{Mazumdar1994}
	D.~Mazumdar, R.~I.~L. Guthrie, {A Comparison of three mathematical modeling
		procedures for simulating fluid flow phenomena in bubble-stirred ladles},
	Metallurgical and Materials Transactions B 25~(2) (1994) 308--312.
	
	\bibitem{Mazumdar1995b}
	D.~Mazumdar, R.~I.~L. Guthrie, {On the numerical computation of turbulent fluid
		flow in CAS steelmaking operations}, App. Math. Model. 19~(9) (1995)
	519--524.
	
	\bibitem{Palovaara2018}
	T.~Palovaara, V.-V. Visuri, T.~Fabritius, Physical modelling of gas injection
	in a ladle, in: Proceedings of the 7th International Congress on Science and
	Technology of Steelmaking, 2018.
	
	\bibitem{Mazumdar1993}
	D.~Mazumdar, R.~I.~L. Guthrie, Y.~Sahai, {On mathematical models and numerical
		solutions of gas stirred ladle systems}, App. Math. Model. 17~(5) (1993)
	255--262.
	
	\bibitem{Castillejos1987}
	A.~H. Castillejos, J.~K. Brimacombe, Measurements of physical characteristics
	of bubbles in gas-liquid plumes: Part ii. local properties of turbulent
	air-water plumes in vertically injected jets, Metall. Trans. 18B (1987)
	595--601.
	
	\bibitem{Castillejos1989}
	A.~H. Castillejos, J.~K. Brimacombe, Physical characteristics of gas jets
	injected vertically upward into liquid metal, Metall. Trans. 20B~(5) (1989)
	595--601.
	
	\bibitem{Joh16}
	V.~John, Finite element methods for incompressible flow problems, Vol.~51 of
	Springer Series in Computational Mathematics, Springer, Cham, 2016.
	
\end{thebibliography}
\end{document}